\documentclass[intlimits,twoside,a4paper]{article}

\usepackage[cp1251]{inputenc}
\usepackage[eqsecnum]{cmpj3}
\usepackage{comment}

\issue{2026}{29}{1}{13402}
\doinumber{10.5488/CMP.29.13402}
\title[Hard-sphere fluids in polydisperse porous media]%
{Thermodynamics of hard-sphere fluids in polydisperse random porous media: {Extended} scaled particle theory%
\thanks{{Dedicated to the 75th anniversary of Prof. Stefan Sokolowski.}}
}
\author[T. Hvozd, M. Hvozd, M. Holovko]{
T. Hvozd\orcid{0000-0002-0156-1753}\thanks{Corresponding author: \email{tarashvozd@gmail.com}.},
M. Hvozd\orcid{0000-0001-8609-309X},
M. Holovko\orcid{0000-0001-8114-5356}\thanks{Email: \email{holovko@icmp.lviv.ua}.}
        }
\address{
Yukhnovskii Institute for Condensed Matter Physics of NAS of Ukraine, 1 Svientsitskii Str., 79011 Lviv, Ukraine
}
\date{Received 15 February 2026; revised 24 February 2026; accepted 24 February 2026; published 30 March 2026}

\begin{document}

\maketitle

\begin{abstract}

Accurate descriptions of reference systems are a central task in liquid-state theories for the study of more complex systems. Using scaled particle theory (SPT), we derive a fully analytical description of the thermodynamic properties of a hard-sphere (HS) fluid confined in size-polydisperse HS random porous media, extending the exis\-ting approaches to higher matrix packing fractions. We calculate chemical potentials for a wide range of porous-matrix parameters, including the matrix packing fraction, degree of polydispersity, and particle-size distributions. Within the proposed framework, our results show excellent agreement with available Monte Carlo simulations and previous integral-equation theories over a broad range of matrix packing fractions, $0.1 \leqslant \eta_0 \leqslant 0.3$, and degrees of polydispersity.

%
%
\end{abstract}

\section{Introduction}


Fluids confined in disordered porous environments exhibit thermodynamic and structural properties that differ significantly from those of bulk systems due to the combined effects of excluded volume, geometrical constraints, and quenched disorder. Such conditions are common in soft matter systems, including colloidal suspensions in porous gels or solids~\cite{Gelb1999}, ionic liquids in nanoconfinement~\cite{Wu2022ChemRev,Kondrat2023ChemRev}, and polymer and protein solutions under macromolecular crowding~\cite{Qin2017,Vlachy2023Biomolecules}, where confinement plays a crucial role in determining the equilibrium behavior and material response.

Hard-sphere (HS) models provide a fundamental reference for understanding these effects, since they isolate purely entropic contributions arising from packing and excluded volume~\cite{Royall2024RMP}. In the context of monodisperse random porous media,  the thermodynamic and structural properties have been studied using a variety of theoretical approaches, including integral equation methods~\cite{Given1992comment,Lomba1993PRE,Given1994,Trokhymchuk1996JPC,Trokhymchuk1997JCP}, density functional theory~\cite{Schmidt2002PRE,Schmidt2005JPCM}, scaled particle theory (SPT)~\cite{Holovko2009JPCB,ChenHolovko2010JPCBcomment,Patsahan2011JCP,Holovko2013PureApplChem,Holovko2015book,ChenHolovko2016JPCB} and computer simulations~\cite{Patsahan2011JCP,Holovko2012CMP,Holovko2017CMP}. 

In porous media, HS fluids are served as reference systems for more complex interactions.
Among these approaches, SPT has proven particularly useful due to its analytical character.
{Recently, original SPT~\cite{Reiss1959JCP,Reiss1960JCP,Lebowitz1965JCP}} has been successfully extended~\cite{Holovko2009JPCB,ChenHolovko2010JPCBcomment,Patsahan2011JCP,Holovko2013PureApplChem,Holovko2015book,ChenHolovko2016JPCB} to describe HS fluids confined in random porous matrices modelled as quenched collections of immobile obstacles {(Madden--Glandt model~\cite{Madden1988JStatPhys,Madden1992JCP})}. Within this framework, analytical expressions for chemical potentials, pressures, and adsorption properties have been obtained for monodisperse fluids in monodisperse porous media, showing good agreement with simulation results at low and intermediate packing fractions of matrix particles. These developments have established SPT as a valuable tool for studying confined fluids in disordered environments. In this way, phase behavior of Lennard-Jones fluid~\cite{Holovko2015CMP}, anisotropic fluids~\cite{HvozdM2018JPCB,Holovko2018CMP,Holovko2020CMP}, ionic liquids~\cite{Holovko2017JMLa,Holovko2017JMLb,HvozdM2022JML,Hvozd2025JML}, polydisperse fluids~\cite{Hvozd2017SM,Hvozd2018JPCB}, patchy colloids~\cite{Kalyuzhnyi2014JPCL,Hvozd2022JCP}, and model antibody fluid~\cite{Hvozd2020SM,Hvozd2022SM} in monodisperse random porous media have been studied.

However, in many soft matter systems of practical interest, size-polydispersity is an intrinsic feature. Colloidal particles, polymers, and biomolecular assemblies typically exhibit broad size distributions, and even moderate polydispersity can strongly affect thermodynamic properties, packing behavior, and phase equilibria. Though the thermodynamics of fluids in one component porous matrices has been extensively investigated, including within SPT-based and related analytical frameworks, considerably less attention has been paid to the fluids under confinement in multicomponent (or polydisperse) porous media.

The existing studies of HS fluids in disordered polydisperse matrices have often relied on numerical calculations and on simulation treatments that restrict polydispersity of the matrix, or assume narrow size distributions~\cite{Ilnytsky1999JPCB,Pizio2000,Rzysko2002JCP,Patsahan2004CMP,Leon2006}. As a result, a fully analytical description that simultaneously accounts for size-polydispersity in the porous matrix remains limited. Such a description is highly desirable, as it would provide transparent insight into how the matrix packing fraction, matrix polydispersity, and various distributions jointly influence the thermodynamic behavior.

In this work, we develop a new approximation within the scaled particle theory description of HS fluids confined in polydisperse random porous media, explicitly incorporating size-polydispersity in the matrix obstacles. Treating the porous medium as a quenched polydisperse HS matrix and the confined fluid as an annealed monodisperse system, we derive closed analytical expressions for the excess chemical potentials and pressures. The proposed approach allows for a systematic analysis of the effects of matrix packing fraction and polydispersity on the thermodynamics of confined HS fluids.

The theoretical predictions are evaluated for representative size distributions and are compared with available Monte Carlo simulation data and established integral equation theoretical results in relevant cases. We find that the present SPT-based framework provides an accurate description of monodisperse HS fluids in polydisperse random porous media, extending the existing analytical theories and offering a useful reference for future studies of confined soft-matter systems.

This paper is dedicated to the 75th anniversary of the birth of our good friend and colleague Stefan Sokolowski. He was one of the leaders in the development of the statistical-mechanical theory of adsorption of fluids and interfacial phenomena. Among his numerous contributions there is a development of the replica Ornstein-Zernike (ROZ) integral equation theory for associative fluids adsorbed in random porous media~\cite{Trokhymchuk1996JPC,Trokhymchuk1997JCP} and for HS fluids adsorbed in polydisperse porous media~\cite{Ilnytsky1999JPCB,Pizio2000,Rzysko2002JCP,Leon2006}. In the present work, we generalize the scaled particle theory SPT2 developed in our group~\cite{Holovko2009JPCB,ChenHolovko2010JPCBcomment,Patsahan2011JCP,Holovko2012CMP,Holovko2017CMP} for HS fluids in a polydisperse HS matrix. As a result, we obtain a series of SPT2b analytical approximations for the thermodynamic properties of HS fluids in polydisperse matrices. We also propose a new, simple approximation, SPT2b3**, within which we achieve an excellent agreement with computer simulation results for HS fluids in polydisperse porous matrices obtained by S. Sokolowski and coworkers~\cite{Leon2006}.

\section{Generalization of the SPT2 approach to HS fluids in size-polydisperse disordered HS matrices}

In contrast to the fully numerical ROZ approach~\cite{Given1992comment,Lomba1993PRE,Given1994,Trokhymchuk1996JPC,Trokhymchuk1997JCP}, the SPT framework provides rather accurate analytical expressions for the chemical potential and pressure of a HS fluid confined in a HS matrix~\cite{Holovko2009JPCB,ChenHolovko2010JPCBcomment,Patsahan2011JCP,Holovko2013PureApplChem,Holovko2015book,Holovko2012CMP,Holovko2017CMP}. The core idea of SPT is to determine the thermodynamic properties of a fluid by analyzing the work required to create a cavity of variable size inside the system. More precisely, SPT relates macroscopic thermodynamic
quantities to the reversible work needed to insert a scaled test particle into the fluid, ranging
from an infinitesimally small particle up to the size of real fluid particles. The exact result for
an infinitesimal scaled particle in a HS fluid confined in a disordered matrix was obtained in~\cite{Holovko2009JPCB}. 
However, the approach developed in~\cite{Holovko2009JPCB}, referred to as SPT1, contained a subtle inconsistency that arises when a 
matrix particle size is significantly larger than a fluid particle size~\cite{ChenHolovko2010JPCBcomment}. This inconsistency was eliminated in~\cite{Patsahan2011JCP} within the framework of a new approach named SPT2. Subsequently, the SPT2 approach for HS fluid in disordered porous media was generalized to HS mixture in multicomponent HS matrix~\cite{ChenHolovko2016JPCB}. This result will be used here for the generalized SPT2 theory for HS fluid in size-polydisperse disordered HS matrix. We note that the reversible work needed to create the cavity by a scaled particle in HS fluid which is free from any other fluid particles is equal to the excess chemical potential of the scaled particle, $\mu_s^{\mathrm{ex}}$. By generalizing our previous exact result for an infinitesimal scaled particle~\cite{Holovko2009JPCB}, the expression for the excess chemical potential of a small scaled particle in the presence of a porous medium can be written in more general form~\cite{Holovko2015book,ChenHolovko2016JPCB}:
\begin{equation}
\label{mu_small_particle}
\beta\mu_s^{\mathrm{ex}}(\lambda_s)=\ln p_0(\lambda_s)-\ln\left[1-\eta_1\frac{(1+\lambda_s)^3}{p_0(\lambda_s)}\right].  
\end{equation}
Here, $\beta=1/k_{\mathrm{B}}T$, where $k_{\mathrm{B}}$ is the Boltzmann constant and $T$ is the temperature; $\lambda_s=R_s/R_1$ is the scaling parameter, with $R_s$ the radius of scaling particle and $R_1=\sigma_1/2$ the radius of a fluid particle ($\sigma_1$~is the diameter). The fluid packing fraction is $\eta_1=\rho_1 V_1$, where $\rho_1$ is the fluid density and $V_1=4\piup R_1^3/3$ is the volume of a fluid particle. We use here conventional notations~\cite{Holovko2009JPCB,Patsahan2011JCP,Holovko2013PureApplChem}, in which the index ``1'' denotes a fluid component, the index ``0'' denotes matrix particles, while for the scaled particle the index ``s'' is used. The term $p_0(\lambda_s)=\exp[-\beta \mu_{s}^0(\lambda_s)]$ is defined by the
excess chemical potential of a scaled particle confined in an empty matrix, $\mu_{s}^0(\lambda_s)$. It represents the probability of finding a cavity created by the scaled particle in the matrix in the absence
of fluid particles.

For a large scaled particle, the excess chemical potential is determined by the thermodynamic expression: 
\begin{equation}
\label{mu_large_particle}
\beta\mu_{s}^{\mathrm{ex}}(\lambda_s)=w(\lambda_s)+ \frac{\beta P V_s}{p_0(\lambda_s)}, 
\end{equation}
where $P$ is the pressure of fluid, and $V_s$ is the volume of the scaled particle. The  multiplier ${V_s}/p_0(\lambda_s)$ arises from the excluded volume occupied by matrix particles and can be interpreted as the probability of finding a cavity created by a scaled particle in the matrix in the absence of fluid particles.
The original SPT2 approach~\cite{Patsahan2011JCP,Holovko2015book} includes two parameters that characterize the porosity of the matrix: the geometric porosity, $\phi_0$, and the thermodynamic porosity, $\phi_1$ (probe particle
porosity). The geometric porosity depends solely on the structure of the matrix and is related to the volume of the void space between the matrix particles,
\begin{equation}
\label{phi_0_with_lambda_M}
\phi_0=p_0(\lambda_s=0).
\end{equation} 
The probe particle porosity is determined by the excess chemical potential, $\mu_1^0$, of fluid particles in the limit of infinite dilution in the matrix and depends on the nature of the fluid under study:
\begin{eqnarray}
\label{probeporosity01_M}
&&\phi_{1}=p_0(\lambda_s=1)=\exp\big(-\beta\mu_1^0\big).
\end{eqnarray}
$w(\lambda_s)$ in (\ref{mu_large_particle}) is the surface term to the $\mu_s^{\mathrm{ex}}(\lambda_s)$ and, according to the ansatz of SPT2 theory~\cite{Holovko2015book} $w(\lambda_s)$~can be presented in the form of expansion:
\begin{equation}
\label{w_expansion}
w(\lambda_s)=w_0+w_1\lambda_s+\frac{1}{2}w_2\lambda_s^2.
\end{equation}
The coefficients $w_0$, $w_1$, and $w_2$ of this expansion  can be determined from the continuity of $\mu_s^{\mathrm{ex}}(\lambda_s)$ and its derivatives, $\partial\mu_s^{\mathrm{ex}}/\partial\lambda_s$ and $\partial^2\mu_s^{\mathrm{ex}}/\partial\lambda_s^2$, at $\lambda_s=0$. By setting $\lambda_s=1$, expression~(\ref{mu_large_particle}) provides the relation between the pressure, $P$, and the excess chemical potential, $\mu_1^{\mathrm{ex}}$, of the fluid in the matrix,
\begin{equation}
\label{mu_ex_book_2015}
\beta\big(\mu_1^{\mathrm{ex}}-\mu_1^0\big)=\ln(1-\eta_1/\phi_0)+A\frac{\eta_1/\phi_0}{1-\eta_1/\phi_0}+B\frac{(\eta_1/\phi_0)^2}{(1-\eta_1/\phi_0)^2}+\frac{\beta P}{\rho_1}\frac{\eta_1}{\phi_1},
\end{equation}
where the coefficients $A$ and $B$ are given by:
\begin{eqnarray}
\label{aBIGp}
&&A=6-4\frac{p'_0}{\phi_0}+\left(\frac{p'_0}{\phi_0}\right)^2-\frac{1}{2}\frac{p''_0}{\phi_0}
\label{bBIGp}, \nonumber \\
&&B=\frac{1}{2}\bigg(3-\frac{p'_0}{\phi_0}\bigg)^2,
\end{eqnarray}
and determine the structure of the disordered porous medium. Here $p'_0=\frac{\partial p_0{\left(\lambda_s\right)}}{\partial \lambda_s}\mid_{\lambda_s=0}$, $p''_0=\frac{\partial^2 p_0{\left(\lambda_s\right)}}{\partial \lambda_s^2}\mid_{\lambda_s=0}$.

Using the Gibbs--Duhem equation,
\begin{equation}
\label{Gibbs_Duhem_eq}
\left(\frac{\partial P}{\partial \rho_1}\right)_T=\rho_1\left(\frac{\partial \mu_1}{\partial \rho_1}\right)_T,
\end{equation}
one can obtain analytical expressions for the chemical potential and pressure of a fluid in porous media within the SPT2 framework. However, the obtained expressions have a defect connected with the divergence at $\eta_1=\phi_1$. To avoid this divergence, it is possible to replace $\phi_1$ with $\phi_0$. Replacing $\phi_1$ with $\phi_0$ everywhere in the SPT2 approach leads to the SPT2a approach in which the chemical potential and the pressure are determined as
\begin{eqnarray}
\label{mu1P_SPT2a_}
\beta\mu_{1}^{\rm{SPT2a}}&=&\ln\big(\Lambda_1^3\rho_1\big)+\beta\mu_1^0-\ln\left(1-\eta_1/\phi_0\right)+\left(1+A\right)\frac{\eta_1/\phi_0}{1-\eta_1/\phi_0}\nonumber \\
&+&\left(\frac{A}{2}+B\right)\left(\frac{\eta_1/\phi_0}{1-\eta_1/\phi_0}\right)^2
+\frac{2B}{3}\left(\frac{\eta_1/\phi_0}{1-\eta_1/\phi_0}\right)^3,
\end{eqnarray}
\begin{eqnarray}
\label{pressureP_SPT2a_}
&&\left(\frac{\beta P}{\rho_1}\right)^{\rm{SPT2a}}=
\frac{1}{1-\eta_1/\phi_0}
+\frac{A}{2}\frac{\eta_1/\phi_0}{\left(1-\eta_1/\phi_0\right)^2}
+\frac{2B}{3}\frac{\left(\eta_1/\phi_0\right)^2}{\left(1-\eta_1/\phi_0\right)^3}.
\end{eqnarray}
Here $\rho_1$ is the density of fluid particles, $\Lambda_1$ is the thermal wavelength and
\begin{equation}
\eta_{1}=\piup\rho_{1}\sigma_{1}^{3}/6
\label{eta}
\end{equation}
is the packing fraction of the fluid particles. However, SPT2a underestimates the influence of porous media. In order to improve the description, it is better to use SPT2b approach which includes the logarithmic divergence at $\eta_1=\phi_1$, 
\begin{eqnarray}
\label{mu1P_SPT2b_}
&&\beta\mu_{1}^{\rm{SPT2b}}=\beta\mu_{1}^{\rm{SPT2a}}-\ln\left(1-\eta_1/\phi_1\right)+\ln\left(1-\eta_1/\phi_0\right).
\end{eqnarray}

The next approximation proposed in~\cite{Holovko2012CMP} is called SPT2b1 and corrects SPT2b by removing the divergence at $\eta_1=\phi_1$ through an expansion of the logarithmic terms in SPT2b,
\begin{eqnarray}
\label{lnSPT2b1_}
&&\ln\left(1-\eta_1/\phi_1\right)\approx\ln\left(1-\eta_1/\phi_0 \right)-\frac{\phi_0-\phi_1}{\phi_1}\frac{\eta_1/\phi_0}{1-\eta_1/\phi_0}.
\end{eqnarray} 
The chemical potential can then be expressed as:
\begin{eqnarray}
\label{mu1P_SPT2b1_}
&&\beta\mu_{1}^{\rm{SPT2b1}}=\beta\mu_{1}^{\rm{SPT2a}}+\frac{\phi_0-\phi_1}{\phi_1}\frac{\eta_1/\phi_0}{1-\eta_1/\phi_0}.
\end{eqnarray}

Other approximations developed in our group~\cite{Holovko2017CMP,Holovko2012CMP} introduce a third type of porosity, $\phi^*$, which represents the maximum allowable packing fraction of fluid particles in the considered matrix. This approximation provides a correct description of the thermodynamic properties of the system across the entire range of low, medium, and high densities. To avoid the divergence $\eta_1=\phi_1$, we follow the algorithm proposed in~\cite{Holovko2017CMP} where the logarithmic term is  expanded in a series around $\left(\phi_1-\phi^\star\right)$~\cite{Holovko2012CMP}, where $\phi^\star$ is defined as 
\begin{eqnarray}
\label{phistar_}
\phi^\star=\frac{\phi_0\phi_1}{\phi_0-\phi_1}\ln\frac{\phi_0}{\phi_1},
\end{eqnarray} 
which is exact for the one-dimensional case~\cite{Reich2004} and can be considered as a good approximation for the three-dimensional case.
The logarithmic term in the SPT2b approach~(\ref{mu1P_SPT2b_}) can be rewritten as
\begin{eqnarray}
\label{lnSPT2b3star_}
&&\ln\left(1-\eta_1/\phi_1\right)\approx\ln\left(1-\eta_1/\phi_0 \right)-\frac{\phi_0-\phi^\star}{\phi^\star}\frac{\eta_1/\phi_0}{1-\eta_1/\phi_0}-\frac{\phi^\star-\phi_1}{\phi^\star}\frac{\eta_1/\phi^\star}{1-\eta_1/\phi^\star}.
\end{eqnarray} 

Taking into account expansion~(\ref{lnSPT2b3star_}) and equation~(\ref{mu1P_SPT2b_}), we obtain expressions for the chemical potential and pressure in the SPT2b3* approximation for the system under study:
\begin{eqnarray}
\label{chemSPT2b3star_}
&&\beta \mu_{1}^{\rm{SPT2b3}^\star}=\beta \mu_{1}^{\rm{SPT2a}}
+\frac{\phi_0-\phi^\star}{\phi^\star}\frac{\eta_1/\phi_0}{1-\eta_1/\phi_0}+\frac{\phi^\star-\phi_1}{\phi^\star}\frac{\eta_1/\phi^\star}{1-\eta_1/\phi^\star},
\end{eqnarray}
\begin{eqnarray}
\label{pressureSPT2b3star_}
\left(\frac{\beta P}{\rho_1}\right)^{\rm{SPT2b3}^\star}&=&
\left(\frac{\beta P}{\rho_1}\right)^{\rm{SPT2a}}
+\frac{\phi_0-\phi^\star}{\phi^\star}\frac{\phi_0}{\eta_1}\left[\ln(1-\eta_1/\phi_0)+\frac{\eta_1/\phi_0}{1-\eta_1/\phi_0}\right] \nonumber \\
&+&\frac{\phi^\star-\phi_1}{\eta_1}\left[\ln(1-\eta_1/\phi^\star)+\frac{\eta_1/\phi^\star}{1-\eta_1/\phi^\star}\right].
\end{eqnarray}

In the next section, we show that SPT2b3* exhibits some problems at higher matrix packing fractions~$\eta_0$, and we introduce a new approximation, SPT2b3**, which can be obtained from~(\ref{chemSPT2b3star_}) by neglecting the last term in the expansion~(\ref{lnSPT2b3star_}). As a result, we obtain:
\begin{eqnarray}
\label{chemSPT2b3star2_}
&&\beta \mu_{1}^{\rm{SPT2b3}^{\star\star}}=\beta \mu_{1}^{\rm{SPT2a}}
+\frac{\phi_0-\phi^\star}{\phi^\star}\frac{\eta_1/\phi_0}{1-\eta_1/\phi_0},
\end{eqnarray}
\begin{eqnarray}
\label{pressureSPT2b3star2_}
&&\left(\frac{\beta P}{\rho_1}\right)^{\rm{SPT2b3}^{**}}=
\left(\frac{\beta P}{\rho_1}\right)^{\rm{SPT2a}}+\frac{\phi_0-\phi^\star}{\phi^\star}\frac{\phi_0}{\eta_1}\left[\ln(1-\eta_1/\phi_0)+\frac{\eta_1/\phi_0}{1-\eta_1/\phi_0}\right].
\end{eqnarray}
Taking into account the Carnahan--Starling correction~\cite{Boublik1975,Yukhnovski2025book,HvozdM2018JPCB}, the expression for the pressure of a HS fluid can be written accordingly:
\begin{eqnarray}
\label{pressureCS1_}
\beta P^{\rm{HS}}=\beta P^{\rm{SPT2...}}+\beta \Delta P^{\rm{CS}},
\end{eqnarray}
where $\beta P^{\rm{SPT2...}}$ is the corresponding SPT2 approximation used in this work (SPT2a, SPT2b1, SPT2b3* or SPT2b3**)
and the second term is the Carnahan--Starling correction:
\begin{eqnarray}
\label{pressCS1_}
\frac{\beta \Delta P^{\rm{CS}}}{\rho_1}=-\frac{\left(\eta_1/\phi_0\right)^3}{\left(1-\eta_1/\phi_0\right)^3}\,.
\end{eqnarray}
Similarly, the expression for the chemical potential of a HS fluid in a size-polydisperse HS porous matrix can be written as
\begin{eqnarray}
\label{muCSandPL_}
\beta\mu_1^{\rm{HS}}=\beta\mu_1^{\rm{SPT2...}}+\beta\Delta\mu_1^{\rm{CS}},
\end{eqnarray}
were the term $\beta \mu_1^{\mathrm{SPT2}\ldots}$ denotes the corresponding SPT2 approximation used in this work (SPT2a, SPT2b1, SPT2b3*, or SPT2b3**), as given by equations~(\ref{mu1P_SPT2a_}), (\ref{mu1P_SPT2b1_}), (\ref{chemSPT2b3star_}), and (\ref{chemSPT2b3star2_}), respectively. The Carnahan--Starling correction to the chemical potential is given by
\begin{eqnarray}
\label{muCS}
&&\beta\Delta\mu_1^{\rm{CS}}=-\frac{\left(\eta_1/\phi_0\right)^3}{\left(1-\eta_1/\phi_0\right)^3}\bigg[\ln(1-\eta_1/\phi_0)+\frac{\eta_1/\phi_0}{1-\eta_1/\phi_0}-\frac{1}{2}\frac{\left(\eta_1/\phi_0\right)^2}{\left(1-\eta_1/\phi_0\right)^2}\bigg].
\end{eqnarray}
In all presented expressions for the chemical potential and the pressure of HS fluids, the influence of porous media is defined by the geometric porosity,
\begin{equation}
\label{phi_0_with_lambda_M_}
\phi_0=p_0(\lambda_s=0),
\end{equation} 
and the thermodynamic porosity,
\begin{eqnarray}
\label{probeporosity01_M_}
&&\phi_{1}=p_0(\lambda_s=1)=\exp\big(-\beta\mu_1^0\big).
\end{eqnarray}
It is important to emphasize that the geometric porosity, $\phi_0$, characterizes the fraction of free volume not occupied by the porous medium. The thermodynamic porosity, $\phi_1$, characterizes the adsorption of a fluid in a porous medium at infinite dilution. In accordance with definition~(\ref{probeporosity01_M_}), it accounts for the thermodynamic insertion of a test fluid particle into the matrix.

For multicomponent HS matrix,
\begin{equation}
\label{p_0_with_lambda_}
p_0(\lambda_s)=1-\frac{\piup}{6}\sum_\alpha\rho_{0\alpha}\left(\sigma_{0\alpha}+\lambda_s\sigma_1\right)^3. 
\end{equation} 
Here $\rho_{0\alpha}$ is the number density of matrix particles of species~$\alpha$ ($\sum_{\alpha}\rho_{0\alpha}=\rho_0$),  $\sigma_{0\alpha}=2R_{0\alpha}$, $\sigma_{1}=2R_{1}$, and $\lambda_s$ is the scaling parameter.

The introduction of the probe particle porosity can be achieved by constructing an approximate form of the chemical potential based on an appropriate equation of state for a mixture.
In the present analysis, the excess chemical potential of the fluid particles in the limit of infinite dilution $\mu_1^0$ relies on the formulation developed by Salacuse and Stell~\cite{Salacuse1982JCP} obtained through the integration of the compressibility equation for a polydisperse HS model.
This result was improved by Leon et al.~\cite{Leon2006} within the framework of Mansoori--Carnahan--Starling--Leland approximation~\cite{Mansoori1971}. As a result, the expression for $\mu_1^0$ for HS fluid at infinite dilution in polydisperse matrix was presented in the following form~\cite{Leon2006}:
\begin{eqnarray}
\label{mu_one_zero_leon2006}
\beta\mu_1^0&=&-\ln(1-\xi_{03})+\sigma_1\frac{3\xi_{02}}{1-\xi_{03}} \nonumber \\
&+&\sigma_1^2\bigg[\frac{3\xi_{01}}{1-\xi_{03}}
+\frac{3\xi_{02}^2}{\xi_{03}}\frac{1}{(1-\xi_{03})^2}
+\frac{3\xi_{02}^2}{\xi_{03}^2}\ln(1-\xi_{03})\bigg] \nonumber \\
&+&\sigma_1^3\bigg[\frac{\piup}{6}\beta P_0+
\frac{\xi_{02}^3}{\xi_{03}^2}\frac{\xi_{03}-2}{1-\xi_{03}}
-\frac{2\xi_{02}^3}{\xi_{03}^2}\ln(1-\xi_{03})\bigg],
\end{eqnarray}
where 
\begin{eqnarray}
\label{pressure_zero_leon2006}
&&\frac{\piup}{6}\beta P_0=\frac{\xi_{00}}{1-\xi_{03}}+\frac{3\xi_{01}\xi_{02}}{(1-\xi_{03})^2}+\frac{3\xi_{02}^3-\xi_{03}\xi_{02}^3}{(1-\xi_{03})^3}, \quad \text{and} \quad \xi_{0l}=\frac{\piup}{6}\sum_{\alpha}\rho_{0\alpha}\sigma_{0\alpha}^l.
\end{eqnarray}
In addition, as noted by Leon et al.~\cite{Leon2006}, the pore size distribution, $v_1$, can be obtained from the thermodynamic porosity and is defined as follows~\cite{Gregg1982}:
\begin{eqnarray}
\label{def_v1}
v_1=-\left(\frac{\partial \phi_1}{\partial \sigma_1}\right).
\end{eqnarray}

\section{Results and discussion}

\begin{figure}[htbp]
\centering\includegraphics[scale=0.9]{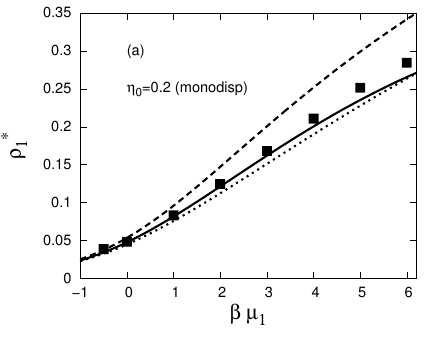}
\centering\includegraphics[scale=0.9]{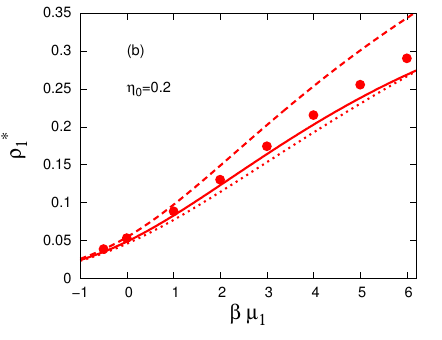}\\
\centering\includegraphics[scale=0.9]{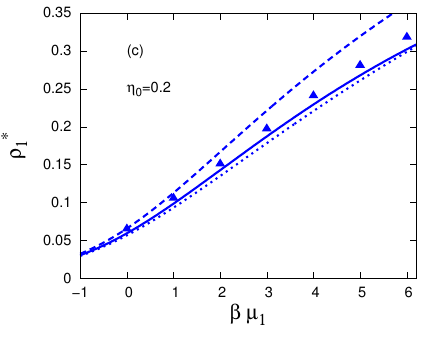}
\centering\includegraphics[scale=0.9]{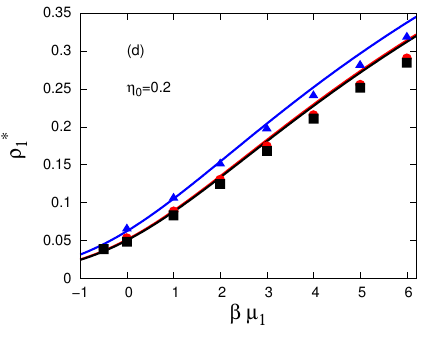}
\caption{(Colour online) {Dependence of the fluid density on the chemical potential}, calculated in this work using the SPT2a (dashed lines), SPT2b1 (dotted lines), and SPT2b3* (solid lines) approximations, and compared against MC simulation data from reference~\cite{Leon2006} (symbols), at fixed matrix packing fraction $\eta_0=0.2$ and for various widths of the rectangular size distribution: (a) $\sigma_0 = 1$, (b) $0.9 \leqslant \sigma_0 \leqslant 1.1$, and (c)~$0.6 \leqslant \sigma_0 \leqslant 1.4$. (d) Same as in panels (a)--(c), but the solid lines correspond to the new SPT2b3** approach.} \label{fig1}
\end{figure}

\begin{figure}[htbp]
\centering\includegraphics[scale=0.9]{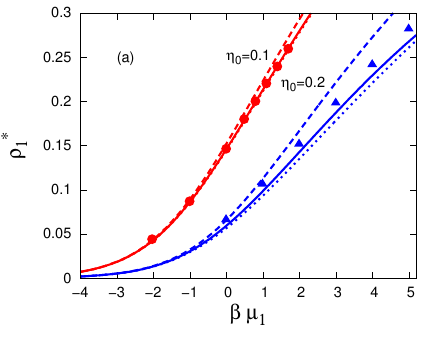}
\centering\includegraphics[scale=0.9]{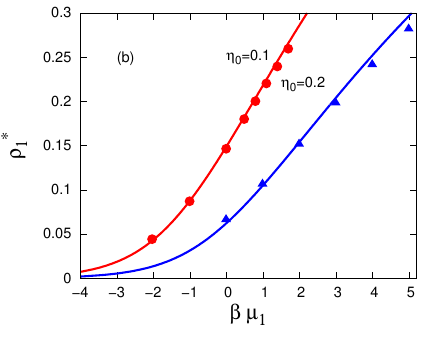}
\caption{(Colour online) {Dependence of the fluid density on the chemical potential}, calculated in this work using (a) the SPT2a (dashed lines), SPT2b1 (dotted lines), and SPT2b3* (solid lines) approximations, and compared against MC simulation data from reference~\cite{Leon2006} (symbols), at fixed width of the rectangular size distribution, $0.6 \leqslant \sigma_0 \leqslant 1.4$, and for various matrix packing fractions: $\eta_0=0.1$ (red lines and filled circles), and $\eta_0=0.2$ (blue lines and filled triangles). (b) Same as in panel (a), but the solid lines correspond to the new SPT2b3** approach.} \label{fig2}
\end{figure}

We consider a monodisperse HS fluid characterized by particle diameter $\sigma_1$ and number density $\rho_1$ (or packing fraction $\eta_1 = \piup \rho_1 \sigma_1^3 / 6$), confined in a porous matrix composed of size-polydisperse HS obstacles quenched in equilibrium. The matrix particles are characterized by a size distribution $f(\sigma_0)$ and number density $\rho_0$ (or packing fraction $\eta_0 = \xi_{03}$).

In the case of the \textbf{size-polydisperse matrix}, the discrete summation is replaced by an integral:
\begin{eqnarray}
\label{xi_zero_l_}
\xi_{0l}=\frac{\piup}{6}\sum_{\alpha}\rho_{0\alpha}\sigma_{0\alpha}^l\longrightarrow \frac{\piup}{6}\int_{0}^{\infty}\rd\sigma_0\rho_0 f(\sigma_0)\sigma_0^l=\frac{\piup}{6}\rho_0 m_l,
\end{eqnarray}
where $m_l$ is the $l$-th moment of polydespersity-size distribution function $f(\sigma_0)$:
\begin{eqnarray}
\label{m_l_}
m_l=\int_{0}^{\infty}\sigma_0^lf(\sigma_0)\rd\sigma_0.
\end{eqnarray}

Taking into account expressions (\ref{p_0_with_lambda_}), (\ref{xi_zero_l_}) and (\ref{m_l_}), in the case of a size-polydisperse matrix we can determine the geometric porosity as follows:
\begin{equation}
\label{p_0_with_moments}
p_0(\lambda_s)=1-\frac{\piup}{6}\rho_0\left[m_3+3\lambda_s\sigma_1 m_2 +3\left(\lambda_s\sigma_1\right)^2 m_1 + \left(\lambda_s\sigma_1\right)^3 m_0\right].
\end{equation}
The presented theoretical expressions in the previous section are sufficient for the investigation of the influence of size-polidespersity on the HS fluid in size-polydisperse porous media. We note that the expressions for $p'_0$ and $p''_0$ needed for calculation $A$ and $B$ in~(\ref{aBIGp}) are:
\begin{eqnarray}
\label{p0_}
p'_0=-\frac{\piup}{2}\rho_0\sigma_1 m_2, \quad p''_0=-\piup\rho_0\sigma_1^2 m_1.
\end{eqnarray}

The influence of the matrix packing fraction, $\eta_0$ (or the geometrical porosity, $\phi_0=1-\eta_0$), for polydisperse matrices with rectangular and Schultz--Zimm size distributions was investigated by Leon~\emph{et~al}.~\cite{Leon2006}. It was shown that, for certain values of the distribution parameters, both distributions lead to very similar values of the thermodynamic porosity $\phi_1$. Moreover, for sufficiently broad size distributions, the thermodynamic porosity, $\phi_1$, is larger than that of a monodisperse matrix at the same geometrical porosity, $\phi_0$, reflecting the more efficient packing of matrix particles of different sizes compared to a one-component HS system.
Another important characteristic of porous materials is the pore size distribution, $v_1$, defined in equation~(\ref{def_v1}). According to the calculations of Leon \emph{et al}.~\cite{Leon2006}, for a rectangular matrix particle size distribution, $v_1$ exhibits a single maximum, the position of which depends on the matrix density. Specifically, increasing the matrix packing fraction shifts this maximum toward smaller test particle diameters. At low matrix packing fractions, $\eta_0$, where the geometrical porosity, $\phi_0$, is close to unity, variations in the matrix particle size distribution have only a minor effect on the pore size distribution,~$v_1$. By contrast, at higher matrix packing fractions (lower $\phi_0$), the position of the maximum of $v_1$ becomes increasingly sensitive to the width of the matrix particle size distribution. Similar qualitative behavior is observed for the Schultz--Zimm distribution.

{Now we calculate the adsorption isotherms.}
To validate our theory and to allow direct comparison with previous studies~\cite{Leon2006}, we first consider a rectangular size distribution:
\begin{equation}
f(\sigma_0)=
\begin{cases}
\dfrac{1}{\sigma_{0,U}-\sigma_{0,L}}, & \sigma_{0,L}\leqslant \sigma_0 \leqslant \sigma_{0,U},\\[6pt]
\;\;\;\;\;\;\;\;0, & \text{otherwise}.
\end{cases}
\end{equation}

In figure~\ref{fig1} we demonstrate the effect of polydispersity on chemical potential at $\eta_0=0.2$ and various widths of the rectangular distribution calculated using different SPT approximations and compared against MC simulations from reference~\cite{Leon2006}. As we can see from figure~\ref{fig1}, increasing matrix size polydispersity decreases the chemical potential at fixed fluid densities. Moreover, the SPT2b3* and SPT2b3** approximations are in good agreement with the MC simulation data, while SPT2b1 and especially SPT2a are less accurate.
However, the simulation data lie between the predictions of the SPT2a and SPT2b1 approximations. As demonstrated in the subsequent figures, this bracketing remains valid at higher values of $\eta_0$. With increasing $\eta_0$, however, the width of this bracket increases, leading to a deterioration in predictive accuracy. At sufficiently large $\eta_0$ and reduced fluid density $\rho_1^*$, the SPT2b3* approximation moves outside this bracket. 
This behavior lies in the core idea for introducing and deriving the new SPT2b3** approximation, which significantly improves the theoretical predictions at higher values of $\eta_0$ and $\rho_1^*$.

\begin{figure}[htbp]
	\centering\centering\includegraphics[scale=0.9]{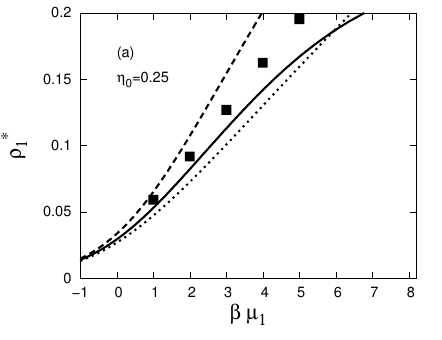}
	\centering\includegraphics[scale=0.9]{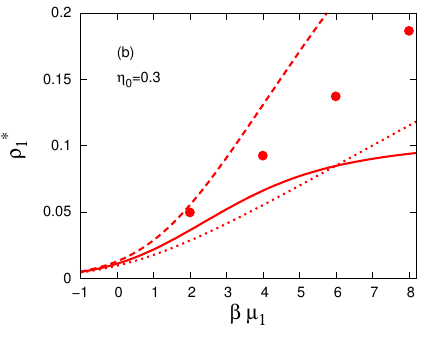}\\
	\centering\includegraphics[scale=0.9]{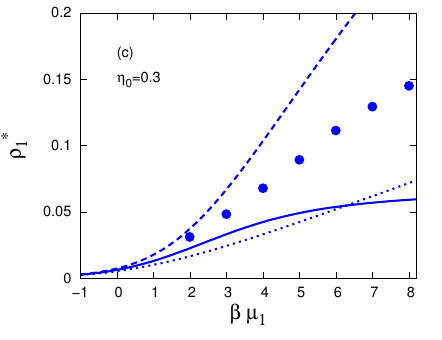}
	\centering\includegraphics[scale=0.9]{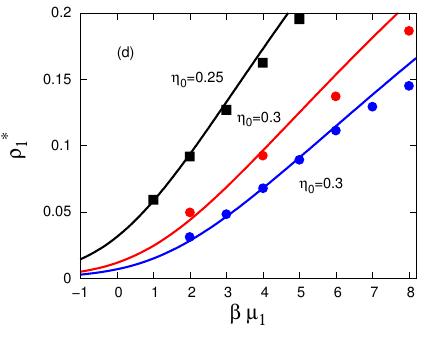}
	\caption{(Colour online) {Dependence of the fluid density on the chemical potential}, calculated in this work using the SPT2a (dashed lines), SPT2b1 (dotted lines), and SPT2b3* (solid lines) approximations, and compared against MC simulation data from reference~\cite{Leon2006} (symbols), for: (a) $\eta_0 = 0.25$ and $0.6 \leqslant \sigma_0 \leqslant 1.4$; (b) $\eta_0 = 0.3$ and $0.6 \leqslant \sigma_0 \leqslant 1.4$; and (c) $\eta_0 = 0.3$ and $0.9 \leqslant \sigma_0 \leqslant 1.1$. (d) Same as in panels (a)--(c), but the solid lines correspond to the new SPT2b3** approach.} \label{fig3}
\end{figure}

Next, in figure~\ref{fig2}, we illustrate the effect of the matrix packing fraction on the chemical potential for a fixed size distribution of matrix particles, $0.6 \leqslant \sigma_0 \leqslant 1.4$. Increasing the matrix packing fraction from $\eta_0 = 0.1$ to $\eta_0 = 0.2$ leads to a significant decrease in the chemical potential at fixed fluid densities. 
It is worth noting that for a comparable matrix packing fraction, $\eta_0 = 0.157$, the case of a monodisperse porous medium was previously studied using SPT and computer simulations~\cite{Holovko2017CMP}. In that work, it was shown that the SPT2b1 and SPT2b3* approximations yield chemical potentials closest to the simulation data. A similar trend is observed in the present study for the size-polydisperse matrix (see figures~\ref{fig1} and~\ref{fig2}).
However, the situation changes drastically with increasing matrix packing fraction. At $\eta_0 = 0.25$, and especially at $\eta_0 = 0.3$, these approximations exhibit noticeable deviations from the simulation results (see figure~\ref{fig3}a, b, and c).
Therefore, as discussed above, we introduce a new approximation, SPT2b3**, which is obtained from equation~(\ref{chemSPT2b3star_}) by neglecting the last term in the expansion given in equation~(\ref{lnSPT2b3star_}). 
In figure~\ref{fig3}(d), we demonstrate excellent agreement between the analytical results obtained within the new SPT2b3** approximation and the MC simulation data for $\eta_0 = 0.25$ (black line and squares) and $\eta_0 = 0.3$ (red line and circles) at a fixed obstacle size distribution, $0.6 \leqslant \sigma_0 \leqslant 1.4$. The same level of agreement is observed for a narrower distribution, $0.9 \leqslant \sigma_0 \leqslant 1.1$, at $\eta_0 = 0.3$ (blue line and circles).

\begin{figure}[h]
	\centering\includegraphics[scale=0.9]{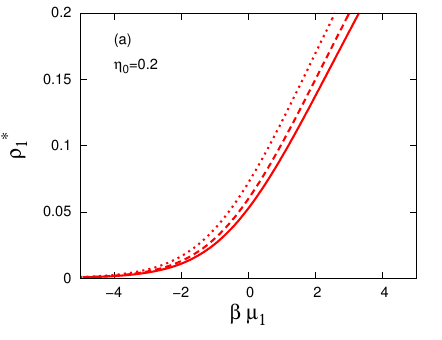}
	\centering\includegraphics[scale=0.9]{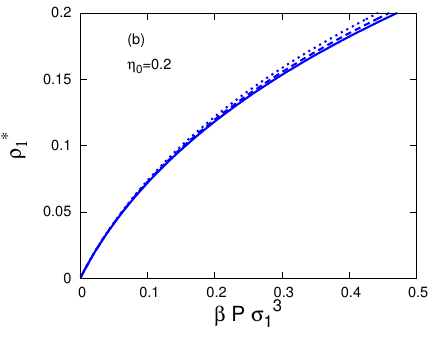}\\
	\centering\includegraphics[scale=0.9]{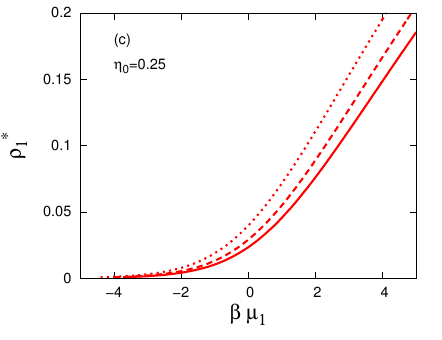}
	\centering\includegraphics[scale=0.9]{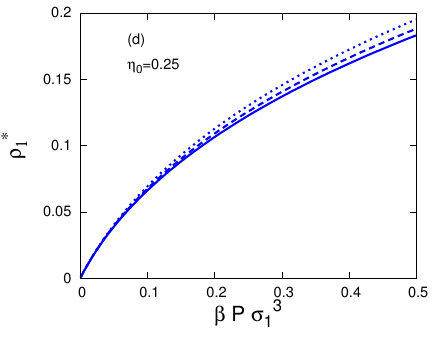}
	\caption{(Colour online) {Dependence of the fluid density on the chemical potential} (a) and on the pressure (b), obtained within the SPT2b3** approximation for a Schultz--Zimm distribution at matrix packing fraction $\eta_0 = 0.2$ and various degrees of polydispersity: $\gamma = 99$ (solid lines), $\gamma = 24$ (dashed lines), and $\gamma = 8$ (dotted lines). Panels (c) and (d) show the corresponding results for $\eta_0 = 0.25$.} \label{fig4}
\end{figure}

Finally, we show that the new SPT2b3** approximation can be simply extended to an  arbitrary continuous obstacle size distribution, which is a key advantage of the proposed SPT-based approach. In particular, in figure~\ref{fig4} we present analytical results for the chemical potentials and pressures when the obstacle matrix is characterized by a Schultz--Zimm size distribution:

\begin{eqnarray}
\label{f_z}
f^{\rm (SZ)}(\sigma_0)=\frac{1}{\gamma !}\left(\frac{\gamma+1}{\bar{\sigma_0}}\right)\sigma_0^\gamma\exp\left[-\left(\frac{\gamma+1}{\bar \sigma_0}\right)\sigma_0\right],
\end{eqnarray}
where
\begin{eqnarray}
\label{f_z}
\int_{0}^{\infty}f^{\rm (SZ)}(\sigma_0)\rd\sigma_0=1,\quad\int_{0}^{\infty}\sigma_0f^{\rm (SZ)}(\sigma_0) \rd\sigma_0=\bar{\sigma_0}.
\end{eqnarray}
 Trends qualitatively similar to those observed for the rectangular distribution are found here. Both the matrix packing fraction and obstacle size polydispersity influence the chemical potential, with the impact of polydispersity becoming increasingly pronounced at higher matrix packing fractions.

\section{Conclusions}

In the present study, we have introduced a new SPT-based approach that provides fully analytical expressions for the thermodynamic properties (the chemical potential and pressure) of a HS fluid confined in size-polydisperse HS random porous matrices. First, we demonstrated excellent agreement with Monte Carlo simulation data for the chemical potentials over a wide range of matrix packing fractions and degrees of polydispersity in the case of a rectangular size distribution. We then applied the approach to systems in which the matrix is characterized by a continuous Schultz--Zimm size distribution and evaluated both chemical potentials and pressures.

The proposed theory extends previous SPT-based descriptions to significantly higher matrix packing fractions and to a broad range of size polydispersities that may be described by arbitrary continuous size distributions, thereby providing a reliable and fully analytical framework for the thermodynamic properties of HS fluids in size-polydisperse random porous media. 

By including additional weak attractive or strong associative interactions, the present model and theoretical approach can serve as a reference system for the description, modelling, and prediction of phase separation, aggregation, and diffusion of protein models in naturally crowded environments~\cite{Hvozd2020SM,Hvozd2022SM,Shah2011JCP,Cho2012PRL,Grimaldo2019JPCL}. 
In particular, expressions for the self-diffusion coefficient of a HS fluid in monodisperse HS porous media obtained in previous studies~\cite{Korvatska2020CMP,Korvatska2025CMP} can be simply extended to polydisperse random porous media by replacing the geometric, $\phi_0$, and thermodynamic, $\phi_1$, porosities with the corresponding expressions derived in the present work.
Furthermore, taking into account the previous studies~\cite{ChenHolovko2016JPCB}, the approach developed here can be generalized to describe size-polydisperse HS fluids confined in size-polydisperse random porous matrices.


\newpage

\bibliographystyle{cmpj}
\bibliography{cmpjxampl_rev}

\ukrainianpart

\title{Термодинаміка твердосферних плинів в полідисперсному невпорядкованому пористому середовищі: узагальнена теорія масштабної частинки}
%
%
\author{Т. Гвоздь, М. Гвоздь, М. Головко}
\address{Інститут фізики конденсованих систем
	імені І.Р. Юхновського НАН України, вул.~Свєнціцького~1, 79011 Львів, Україна}

\makeukrtitle

\begin{abstract}
	\tolerance=3000%
	Точний опис базисних систем є головним завданням теорій рідкого стану для дослідження більш складних систем. Використовуючи теорію масштабної частинки, ми отримали повністю аналітичний опис термодинамічних властивостей плину твердих сфер в невпорядкованому пористому середовищі, яке представлене полідисперсними за розміром твердими сферичними перешкодами. Такий опис узагальнює існуючі підходи на випадок вищих значень коефіцієнта упаковки частинок матриці. Ми розрахували хімічні потенціали для широкого діапазону параметрів пористої матриці, включаючи коефіцієнт упаковки частинок матриці, ступінь полідисперсності та розподіли розмірів частинок матриці. У рамках запропонованого підходу наші результати демонструють відмінну узгодженість із наявними даними комп'ютерного моделювання методом Монте-Карло та теоріями інтегральних рівнянь в широкому діапазоні значень коефіцієнта упаковки частинок матриці, $0.1 \leqslant \eta_0 \leqslant 0.3$, і ступенів полідисперсності.
	\keywords полідисперсність, твердосферний плин, невпорядковане пористе середовище, термодинаміка
	
\end{abstract}

\end{document}